\documentstyle[emulateapj,onecolfloat,psfig]{article}

\newlength{\tskip}
\newlength{\colwidth}

\newcommand{\beq}{\begin{equation}}
\newcommand{\eeq}{\end{equation}}
\newcommand{\beqa}{\begin{eqnarray}}
\newcommand{\eeqa}{\end{eqnarray}}
\begin{document}
\twocolumn[
\title{Weak Lensing by Large-Scale Structure: A Dark Matter Halo
	Approach}
\author{Asantha Cooray$^1$, Wayne Hu$^2$\altaffilmark{4}, and Jordi
Miralda-Escud\'e$^{3,}$\altaffilmark{4}}
\affil{
$^1$Department of Astronomy and Astrophysics, University of Chicago,
Chicago IL 60637\\
$^2$Institute for Advanced Study, Princeton, NJ 08540\\
$^3$Department of Astronomy, McPherson Labs., Ohio State University,
Columbus, OH 43210\\
E-mail: asante@hyde.uchicago.edu, whu@ias.edu, jordi@astronomy.ohio-state.edu}
\submitted{Submitted for publication in The Astrophysical Journal Letters}

%------------------------------------------------------------------------------

\begin{abstract}
Weak gravitational lensing observations probe the spectrum
and evolution of density fluctuations and the cosmological parameters
which govern them but are currently limited to small fields
and subject to selection biases. We show how the expected 
signal from large-scale structure arises from the contributions 
from and correlations between individual halos. 
We determine the convergence power spectrum as 
a function of the maximum halo mass and so provide the means
to interpret results from surveys that lack high mass halos
either through selection criteria or small fields. 
Since shot noise from rare massive halos is mainly responsible 
for the sample variance below 10$'$, our method should aid
our ability to extract cosmological information from small fields. 
\end{abstract}
%------------------------------------------------------------------------------
% User-supplied List of keywords.

]
%------------------------------------------------------------------------------

\altaffiltext{4}{Alfred P. Sloan Fellow}

\section{Introduction}

Weak gravitational lensing of faint galaxies probes the 
distribution of matter along the line of sight.  Lensing by
large-scale structure (LSS) induces
correlation in the galaxy ellipticities at the percent level
(e.g., \cite{Mir91} 1991; \cite{Blaetal91} 1991;
\cite{Kai92} 1992).  Though challenging to measure, these 
correlations
provide important cosmological information that is
complementary to that supplied by
the cosmic microwave background and potentially as precise
(e.g., \cite{JaiSel97} 1997;
\cite{Beretal97} 1997; \cite{Kai98} 1998; \cite{Schetal98}
1998; \cite{HuTeg99} 1999; \cite{Coo99} 1999; \cite{Vanetal99} 1999;
see \cite{BarSch00} 2000 for a recent review).
Indeed several recent studies have provided the first clear evidence
for weak lensing in so-called blank fields (e.g., \cite{Vanetal00} 2000;
\cite{Bacetal00} 2000; \cite{Witetal00} 2000).

Weak lensing surveys are currently limited to small fields which may
not be representative of the universe as a whole, owing to sample
variance.  In particular, rare massive objects can contribute
strongly to the mean power in the shear or convergence but not
be present in the observed fields.  The problem is compounded
if one chooses blank fields subject to the condition that they do not
contain known clusters of galaxies. Our objective in this {\it Letter}
is to
quantify these effects and to understand what fraction of the
total convergence power spectrum should arise from lensing by individual
massive clusters as a function of scale.

In the context of standard cold dark matter (CDM) models for
structure formation, the dark matter halos that are responsible for lensing 
have properties that have been intensely studied by numerical 
simulations.  In particular, analytic scalings and fits now exist
for the abundance, profile, and correlations of halos of 
a given mass.
We show how the convergence power spectrum predicted in these
models can be constructed from these halo properties. 
The critical ingredients
are: the Press-Schechter formalism (PS;
\cite{PreSch74} 1974) for the mass function; the NFW
profile of \cite{Navetal96} (1996), and the halo bias
model of \cite{MoWhi96} (1996).
Following \cite{Sel00} (2000),  
we modify halo profile parameters, specifically concentration, so 
that halos account for the full non-linear dark matter 
power spectrum and generalize his treatment to be applicable 
through all redshifts relevant to current galaxy ellipticity 
measurements of LSS lensing. 
This calculational method allows us to determine the contributions
to the convergence power spectrum of halos
of a given mass.

Throughout this paper, we will take $\Lambda$CDM as our fiducial cosmology 
with parameters $\Omega_c=0.30$ for the CDM density,
$\Omega_b=0.05$ for the baryon density, $\Omega_\Lambda=0.65$ for the
cosmological constant, $h=0.65$ for the dimensionless Hubble
constant and a scale invariant spectrum of
primordial fluctuations, normalized to  galaxy  cluster abundances 
($\sigma_8=0.9$ see \cite{ViaLid99} 1999)  
and consistent with COBE (\cite{BunWhi97} 1997).
For the linear power spectrum, we take the
fitting  formula for the transfer function given in \cite{EisHu99} (1999). 

\begin{figure*}[t]
\centerline{\psfig{file=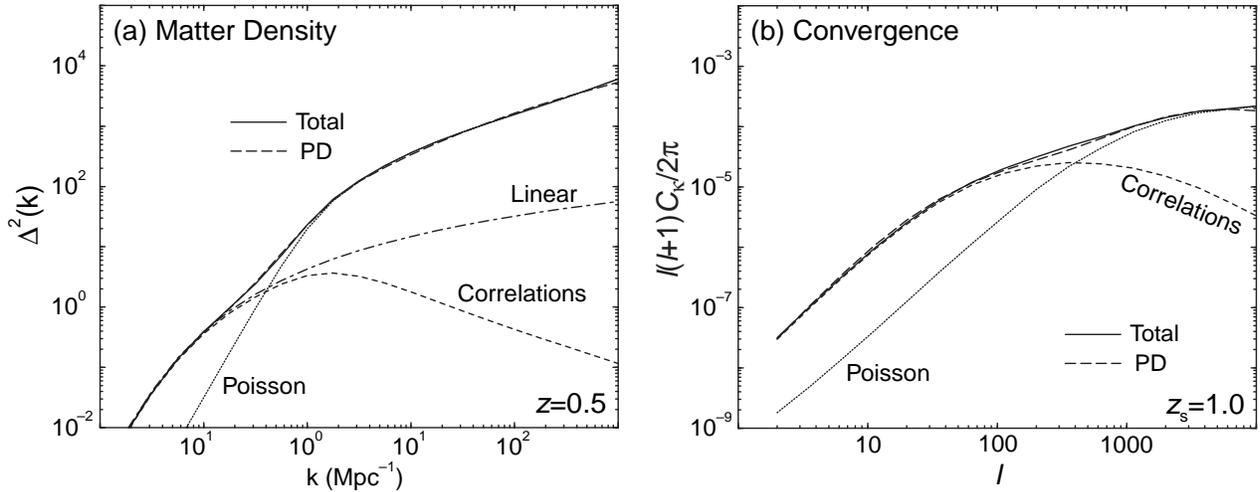,width=0.9\textwidth}}
\caption{Halo model power spectra. (a) Dark matter power spectrum 
at redshift of 0.5. (b) Convergence power spectrum with $z_{s}=1$.
The sum of the Poisson (dotted line) and correlation 
(dashed) contribution (solid line) compares well with
that predicted by the non-linear power spectrum based on
the PD fitting function (long-dashed line). In (a), the linear matter
density power spectrum is shown with a dot-dashed line.}
\label{fig:pd}
\end{figure*}

\section{Lensing by Halos}
\label{sec:model}

\subsection{Halo Profile}

We model dark matter halos as NFW profiles with a density distribution
\begin{equation}
\rho(r) = \frac{\rho_s}{(r/r_s)(1+r/r_s)^{2}} \, .
\end{equation}
The density profile can be integrated and related to the total dark
matter mass of the halo within $r_v$
\begin{equation}
M =  4 \pi \rho_s r_s^3 \left[ \log(1+c) - \frac{c}{1+c}\right] \, ,
\label{eqn:mass}
\end{equation}
where the concentration, $c$, is defined as $r_v/r_s$.
Choosing $r_v$ as the virial radius of the halo, spherical
collapse tells us that 
$M = 4 \pi r_v^3 \Delta(z) \rho_b/3$, where $\Delta(z)$ is
the overdensity of collapse (see
e.g. \cite{Hen00} 2000) and $\rho_b$ is the background matter density
today. We use
comoving coordinates throughout.
By equating these two expressions, one can
eliminate $\rho_s$ and describe the halo by its mass $M$ and 
concentration $c$. 
Finally, we can determine
a relation between $M$ and $c$ such that halo distribution produces the
same power as the non-linear dark matter power spectrum, as outlined
in \cite{Sel00} (2000).

\subsection{Convergence Power Spectrum}

For lensing convergence, we need the projected surface mass density, 
which is the line-of-sight integral of the profile
\begin{equation}
\Sigma (r_\perp) = \int_{-r_v}^{+r_v} \rho(r_\perp,r_\parallel) d r_\parallel \, ,
\end{equation}
where $r_\parallel$ is the line-of-sight distance and $r_\perp$ 
is the perpendicular distance. As in equation~(\ref{eqn:mass}), 
the cut off here at the virial radius 
reflects the fact that we only account for mass contributions
out to $r_v$ (see \cite{Bar96} 1996 for an analytical description
when $r_v \rightarrow \infty$). 
The convergence on the sky $\kappa(\theta)$ is related
to surface mass density through
\begin{equation}
\kappa(\theta) = \left( {4 \pi G \over c^2} {d_l d_{ls} \over d_s}
\right) (1+z_l) \Sigma(d_l \theta) \, ,
\end{equation}
where the extra factor of $(1+z_l)$ from the familiar expression
comes from the
use of comoving coordinates to define densities and distances,
e.g. $d_l$, $d_s$ and $d_{ls}$ are the {\it comoving}
angular diameter distances 
from
the observer to lens, observer to source, and
the lens to source, respectively.

The total convergence power spectrum due to halos, $C_\kappa^{\rm tot}$, can be
split into two parts: a Poisson term, $C_\kappa^{\rm
P}$, and a term involving correlations between individual halos,
$C_\kappa^{\rm C}$.  This split was introduced by
\cite{ColKai88} (1988) to examine the power spectrum of
the Sunyaev-Zel'dovich (SZ; \cite{SunZel80}
1980) effect due to galaxy clusters (see \cite{KomKit99} 1999 and references therein for more recent applications).

The Poisson term due to individual halo contributions can be 
written as,
\begin{equation}
C_\kappa^P(l) = \int_0^{z_{s}} dz \frac{d^2V}{dz d\Omega}
\int_{M_{min}}^{M_{max}} dM \frac{dn(M,z)}{dM} \left[ \kappa_l(M,z)
\right]^2 \, .
\label{eqn:poisson}
\end{equation}
where $z_s$ is the redshift of background sources, $d^2 V/dz d\Omega$ 
is the comoving differential volume, and
\begin{equation}
\kappa_l = 2\pi \int_0^{\theta_v} \theta d\theta \,
\kappa(\theta) J_0\left[\left(l+\frac{1}{2}\right) \theta \right] \, ,
\end{equation}
is the 2D Fourier transform of the halo profile.
The halo mass distribution
as a function of redshift [$dn(M,z)/dM$] is determined through the
PS formalism.  

Here, we have assumed that all sources are at a single redshift; 
for a distribution of sources one 
integrates over the normalized background source
redshift distribution. The minimum, $M_{\rm min}$, and maximum, 
$M_{\rm max}$,
masses can be varied to study the effects of
rare and excluded high mass halos.

The clustering term arises from correlations between halos of
different masses. By assuming that the linear matter density power
spectrum, $P(k,z)$,
is related to the power spectrum of halos over the
whole mass range via a redshift-dependent linear
bias term, $b(M,z)$, we can write the correlation term as
\begin{eqnarray}
C_\kappa^C(l) &=& \int_0^{z_{s}} dz \frac{d^2 V}{dzd\Omega}
P\left(\frac{l}{d_l},z\right) \nonumber \\
&\times& 
\left[ \int_{M_{min}}^{M_{max}} dM \frac{dn(M,z)}{dM} b(M,z) \kappa_l(M,z)
\right]^2 \, .
\label{eqn:corr}
\end{eqnarray}
Here we have utilized the Limber approximation (\cite{Lim54} 1954) by
setting $k = l/d_l$. 
\cite{MoWhi96} (1996) find that the halo bias can be
described by 
$b(M,z) = 1 + [\nu^2(M,z) -1]/\delta_c$, where $\nu(M,z) =
\delta_c/\sigma(M,z)$ is the peak-height threshold, 
$\sigma(M,z)$ is the rms fluctuation within a top-hat filter  at the 
virial radius corresponding to mass $M$,
and $\delta_c$ is the threshold overdensity of spherical
collapse (see \cite{Hen00} 2000 for useful fitting functions).

\begin{figure*}[t]
\centerline{\psfig{file=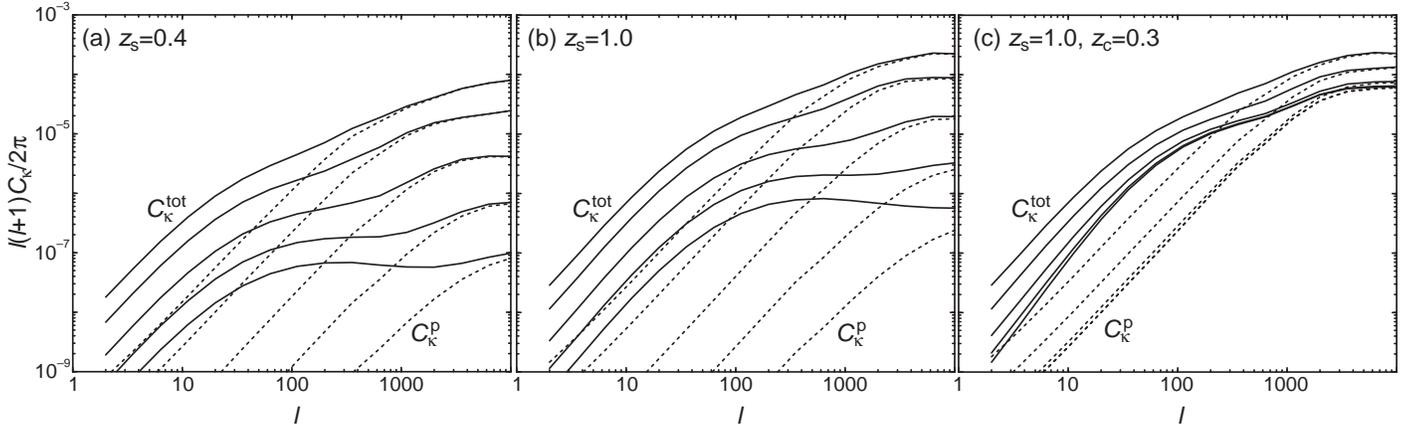,width=\textwidth}}
\caption{Lensing convergence as a function of 
maximum halo mass.  
In the increasing order, for both Poisson (dotted lines)
and total contributions (solid lines), the maximum mass is $10^{11}$, $10^{12}$, $10^{13}$,
$10^{14}$ and $10^{15}$ $M_{\sun}$.  (a) sources at $z_s=0.4$;
(b) $z_{s}=1$; (c) $z_{s}=1$ with mass cut off applied
only out to $z=0.3$.} 
\label{fig:mass}
\end{figure*}

\section{Results}

Following the approach given in \cite{Sel00} (2000), we first
test the halo prescription against the full non-linear density power
spectrum found in simulations and fit by \cite{PeaDod96} (PD, 1996).
In Fig.~\ref{fig:pd}a, as an example,
 we show the comparison at $z=0.5$.
A good match between the two power spectra was achieved by slightly
modifying the concentration relation of \cite{Sel00} (2000) as
\begin{equation}
c(M,z) = a(z) \left[\frac{M}{M_{\star}(z)}\right]^{-b(z)} \, .
\end{equation}
Here, $M_{\star}(z)$ is the non-linear mass scale at which $\nu(M,z)
=1$, while $a(z)$ and $b(z)$ can be considered as adjustable 
parameters. The dark matter power spectrum is well reproduced, to
within 20\% for $0.0001 < k < 500$ Mpc$^{-1}$, out to a redshift of 1
with the parameters $a(z)=10.3(1+z)^{-0.3}$, and
$b(z)=0.24(1+z)^{-0.3}$,  which 
agree with the values given by Seljak (2000) for the NFW profile at $z=0$. 
The two power spectra differ increasingly with scale at $k > 500$ Mpc$^{-1}$, 
but the Peacock and Dodds (1996) power spectrum is not reliable there due to 
the resolution limit of the simulations from which the non-linear power 
spectrum was derived. Note that the above $c(M,z)$ relation is only valid for 
the cosmology used here and for the NFW profile; moreover, it should not
necessarily give the true mean density profile of halos, since other
effects not considered in our halo prescription, such as
halo substructure, would affect the relation between the dark matter power
spectrum and the spatial distribution and mean density profiles of 
halos. A detailed study of $c(M,z)$ as generally applied to all
cosmologies, profile shapes and power spectra is currently in
progress (Seljak, private communication).

In general, the behavior of dark matter power spectrum due to halos
can be understood in the following way. 
The linear portion of the dark matter power spectrum, $k < 0.1$
Mpc$^{-1}$, results from the correlation 
between individual dark matter halos and reflects the
bias prescription.  The fitting formulae of \cite{MoWhi96} (1996) 
adequately describes this regime for all redshifts.
The mid portion of the power spectrum, around $k \sim 0.1-1$
Mpc$^{-1}$ corresponds to the non-linear scale
$M \sim M_{\star}(z)$, where the Poisson and correlated 
term contribute
comparably. At higher $k$'s, the power arises mainly from
the contributions of individual halos (see \cite{Sel00} 2000 for a
discussion of the detailed properties of 
the density  and galaxy power spectra due to halos).

In Fig.~\ref{fig:pd}b, we show the same comparison for
the convergence power spectrum.
The LSS power spectrum was calculated following 
\cite{HuTeg99} (1999) using the \cite{PeaDod96} (1996) power
spectrum for the underlying mass distribution and using the same
Limber approximation as the correlation calculation presented here.
The lensing power spectrum due to halos has the same behavior as the
dark matter power spectrum. At large angles ($l \lesssim 100$),
the correlations between halos dominate. 
The transition from linear to non-linear is at $l \sim
500$ where halos of mass similar to $M_{\star}(z)$ contribute. 
The Poisson contributions start dominating at $l > 1000$.

In order to establish the extent to which massive halos contribute,
we varied the maximum mass of  halos, $M_{\rm max}$, in the convergence
calculation. The results are shown in Fig.~\ref{fig:mass}. We
use a background source redshift of 1 and 0.4, corresponding to deep
lensing surveys and to a shallower survey such as
the ongoing 
Sloan Digital Sky Survey\footnote{http://www.sdss.org}.  
In Fig.~\ref{fig:mass}a,b we
exclude masses above a certain threshold at all redshifts and
in c only for halos below redshift $z=0.3$,
reflecting the fact that current 
observations of galaxy clusters are likely to be complete
only out to such a low redshift. Assuming the latter, 
we find that  a significant contribution
comes from massive clusters at low redshifts (see
Figs.~\ref{fig:mass}b \& c). Ignoring such masses,
say above $\sim$ 10$^{14}$ M$_{\sun}$ can lead to a convergence power
spectrum which is a factor of $\sim$ 2 lower than the total.
Note that such a high mass cut off affects the Poisson
contribution of halos more than the correlated contributions
and can bias the shape not just the amplitude of the power
spectrum. 

\begin{figure}[t]
\centerline{\psfig{file=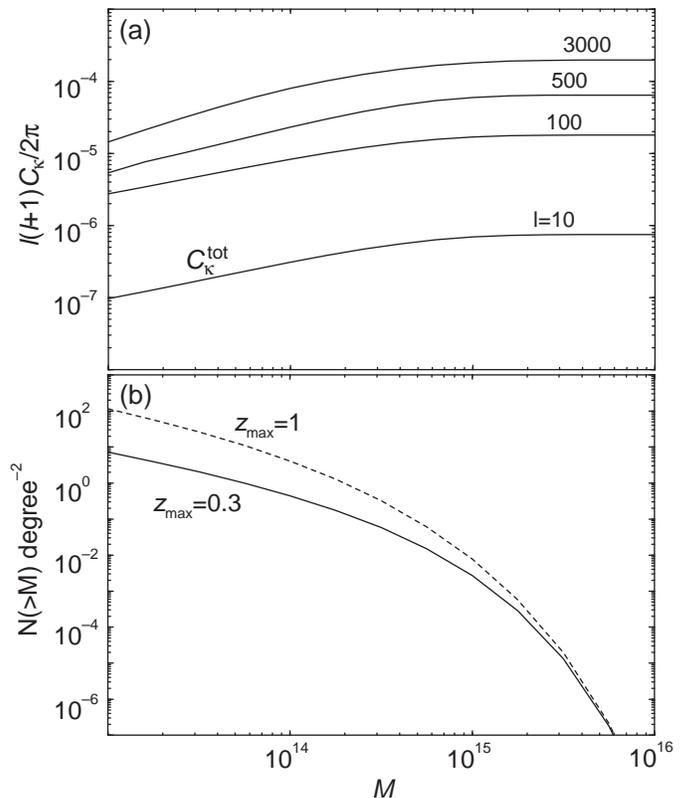,width=3.5truein}}
\caption{(a)
Total lensing convergence $C_\kappa^{\rm tot}$ as a function of
maximum mass for several $l$-values and sources at $z_{s}=1$.
As shown, contributions from halos with masses $>10^{15}$ $M_{\sun}$ 
are negligible.
(b) Surface density of halo masses as a function of minimum
mass using PS formalism out to $z_{\rm max}=0.3$ and $z_{\rm max}=1$. This determines
the survey area needed to ensure a fair sample of halos greater than
a given mass.}
\label{fig:press}
\end{figure}

In Fig.~\ref{fig:press}a, we show the dependence of $C_{\kappa}^{\rm 
tot}$, for several $l$ values.  If halos 
$<10^{15}$ $M_{\sun}$ are well represented in a survey, then
the power spectrum will track the LSS convergence power spectrum
for all $l$ values of
interest.  The surface number density of halos determines
how large a survey should be to possess a fair sample of halos
of a given mass.  We show this in Fig.~\ref{fig:press}b
as predicted by PS formalism for our fiducial
cosmological model for halos out to 
($z=0.3$ and $z=1.0$). Since the surface number density of
$>10^{15} M_{\sun}$ halos out to a redshift of 0.3 and 1.0 is
$\sim$ 0.03 and 0.08 degree$^{-2}$ 
respectively, a survey of  order
$\sim$ 30 degree$^2$ should be sufficient to contain a fair
sample of the universe for recovery of the full LSS convergence
power spectrum.

One caveat is that mass cuts may affect the higher moments
of the convergence differently so that a fair sample for
a quantity such as skewness will require a different survey
strategy.  From numerical simulations
(\cite{WhiHu99} 1999), we know that $S_{3}\equiv \left< \kappa^{3}
\right> /\left< \kappa^{2}\right>^{2}$ shows substantial sample
variance, implying that it may be dominated by 
rare massive halos. When calculated with
hyper-extended perturbation theory (HEPT, see \cite{Hui99}
1999) but 
using the halo power spectrum instead of non-linear density
power spectrum, the skewness decreased by 
a factor of $\sim$ 2.5 to 3 with a mass cut off at
$\sim$ $10^{13}$ M$_{\sun}$. Since it is unclear to what
extent HEPT ansatz remains
valid for the halo description, these results should be taken
as provisional and will be the subject of future study.

While upcoming wide-field weak lensing surveys, such as the MEGACAM
experiment at Canada-France-Hawaii Telescope (\cite{Bouetal98} 1998), and the
proposed wide field survey by Tyson et al. (2000, private
communication) will cover areas up to $\sim$ 30 degree$^2$ or more, 
the surveys that have been so far published, e.g.,
\cite{Witetal00} (2000), only cover at most 4 degree$^2$ in areas
without known clusters. The observed convergence
in these fields should be biased low compared with the mean
and vary widely from field to field due to sample variance from
the Poisson contribution of the largest mass halos in the fields,
which are mainly responsible for the sample variance below $10'$ (see
\cite{WhiHu99} 1999).

Our results can also be used proactively.
If properties of the mass distribution such as the maximum mass
halo in the observed lensing fields are known, say through prior optical,
X-ray, SZ or even internally in the lensing observations (see
\cite{KruSch99} 1999), one can 
make a fair comparison of the observations to theoretical model 
predictions
with a mass cut off in our formalism.  Even for larger surveys,
the identification and extraction of large halo contributions can
be beneficial: most of the sample variance in the fields will be
due to rare massive halos.  A reduction in the sample variance
increases the precision with which the power spectrum
can be measured and hence the cosmological parameters upon which
it depends.

\acknowledgments
We acknowledge useful discussions with  Uros Seljak. WH is supported
by the Keck Foundation and NSF-9513835.

\end{document}